# A test for the existence of isomorphs in glass-forming materials

D. Fragiadakis and C.M. Roland

*Chemistry Division, Naval Research Laboratory, Washington D.C. 20375-5342, USA*

We describe a method to determine whether a material has isomorphs in its thermodynamic phase diagram. Isomorphs are state points for which various properties are invariant in reduced units. Such materials are commonly identified from strong correlation between thermal fluctuations of the potential energy, U, and the virial W, but this identification is not generally applicable to real materials. We show from molecular dynamic simulations of atomic, molecular, and polymeric materials that systems with strong U-W correlation cannot be pressure densified; that is, the density obtained on cooling to the glassy state and releasing the pressure is independent of the pressure applied during cooling.

## INTRODUCTION

Efforts to "solve" the glass transition problem are confronted with the myriad behaviors exhibited by vitrifying liquids and polymers. Theories usually make predictions for the divergence of the primary relaxation time, $\tau_\alpha$, with decreasing temperature, since this is the defining characteristic of glass formation. However, there are numerous other phenomena that must be identified and ultimately addressed by a comprehensive theoretical model. A major development along these lines was the discovery of isomorphs [1,2,3], curves in the phase diagram for which state points having microscopic configurations $\left(r_1^{(1)},\ldots,r_N^{(1)}\right)$ and $\left(r_1^{(2)},\ldots,r_N^{(2)}\right)$ with the same reduced coordinates $\tilde{r}_i \equiv \rho^{1/3} r_i$ ($\rho$ is the density) have proportional canonical probability factors:

$$\exp\left[-U\left(r_1^{(1)},\ldots,r_N^{(1)}\right)/k_B T_1\right] \propto \exp\left[-U\left(r_1^{(2)},\ldots,r_N^{(2)}\right)/k_B T_2\right] \quad (1)$$

where $U$ is the potential energy, $k_B$ the Boltzmann constant, and the proportionality constant depends only on the state points (1) and (2). Isomorphic state points are characterized by certain properties: constancy of $\tau_\alpha$ (a property known as density scaling [4]), isochronal superpositioning of the relaxation dispersion (i.e., invariance of the shape of the relaxation dispersion) [5,6], invariance of thermodynamic properties such as the excess entropy and isochoric specific heat [3], and strong correlations in the equilibrium fluctuations of $U$ and the virial pressure, $W$ [7]. While the first inspired the development of isomorph theory, the last is the property of choice to test whether a material has isomorphs. Additional properties associated with isomorphic state points include a Prigogine-Defay ratio (connecting the changes in thermal expansivity, isothermal compressibility, and heat capacity upon vitrification) having a value near unity [8,9] and simplified physical aging behavior [1]. This class of materials, which includes hypothetical liquids having repulsive, inverse power-law intermolecular potentials [10,11], are referred to as "strongly correlating liquids" [2], "simple liquids" [12,13], or in homage to the group responsible for the isomorph ansatz, "Roskilde liquids" [14]. Herein we adopt the latter to refer specifically to materials having isomorphs in their phase diagram.

The principal drawback to application of the isomorph theory is the difficulty of testing it for real materials. Key quantities such as the microscopic configurational energies or the magnitude of fluctuations of $W$ and $U$ are obtained only through molecular dynamics (MD) simulations. Conformance to density scaling [4] and isochronal superpositioning [5,6] can be assessed experimentally, but the usual measurements encompass only a limited range of thermodynamic conditions, so the conclusions can be tentative. An example is sorbitol, which conforms to density scaling over a limited range of $T$ and $P$ [15], despite lacking isomorphs because of its hydrogen bonding [16]. More generally, MD simulations indicate that strongly polar liquids can exhibit density scaling yet have poor W-U correlation [13]. The proximity of the Prigogine-Defay ratio to a value of unity is a criterion for Roskilde simple behavior; however, determination of this ratio is difficult, requiring measurements of several frequency-dependent thermoviscoelastic response functions [8].

Another means to assess the isomorph theory is from its prediction that $\tau_\alpha$ and the viscosity are constant along the melting line [1,17]. (This statement is strictly true only for properties expressed in reduced units, although the difference between reduced versus actual units is negligible in the supercooled regime [18].) Since the prediction of constant viscosity is only for equilibrium melting, it cannot be tested for polymers or any material in which crystallization is sensitive to thermal history. An evaluation of 43 simple liquids for which melting temperatures, viscosities, and the equation of state were available revealed that 8 qualified as Roskilde liquids; specifically, only for substances with a rigid, spherical shape and no polar bonds was the melting line an isomorph [19].

Given the appeal of identifying a fundamental property that connects prominent characteristics of many glass-forming materials, there is obvious value in bridging the underlying theory and measurements on actual materials. Herein we describe a general method to determine experimentally whether a glass-forming material has isomorphs, as defined by eq. (1). To do this we take advantage of two properties of Roskilde liquids [1,20]: the fact that state points with equal relaxation time (in particular $\tau_\alpha$ at the glass transition) are isomorphic and have identical structure, and the fact that a jump from two isomorphic state points to a third state point results in equivalent aging behavior (see Appendix). We test our idea using MD simulations, and then briefly review the limited results for actual liquids.

The particular procedure we employ is known as pressure densification [21,22,23,24,25,26,27,28,29]. Whereas conventionally glass is formed by quenching at ambient pressure, pressure densification involves application of pressure to the liquid prior to cooling below $T_g$. The pressure is then released, and the material evolves toward equilibrium from the same temperature and pressure as the conventional glass. For a Roskilde simple liquid, the glass transition temperature $T_g(P)$, defines an isomorph. The structure of the glasses formed at ambient pressure ("normal glass") and high pressure ("pressure densified glass") will therefore be identical at $T_g$, and assuming that a similar amount of physical aging occurs for the two glasses during subsequent cooling (this assumption, supported by our simulations, is justified below), after releasing the pressure, the pressure densified glass will be identical to the conventional glass. For a non-Roskilde liquid, which lacks isomorphs, the structure of the glass formed at high pressure will be different than that of the normal glass, and the two will have different properties.

## METHOD

Simulations were carried out using the RUMD simulation software [30], all performed in the NVT ensemble with a Nose-Hoover thermostat [31] or the NPT ensemble using an added Berendsen barostat [32]. To produce the normal glass, the system was cooled at constant rate from well above the glass transition at constant pressure $P_0$=1 to a temperature $T_B$ well below the glass transition (point G "normal glass" in Fig. 1). The cooling rate was in the range of 2–8 × $10^{-5}$ (Lennard-Jones units). $T_B$ is chosen sufficiently low that aging at that temperature is negligible at the timescale of our simulation. The glass transition was identified from a change in slope of the specific volume vs. temperature curve. To produce the pressure densified glass, the sample was first equilibrated well above the glass transition at a pressure, $P_1 > P_0$, then cooled to $T_B$ at the same cooling rate as the normal glass, with the pressure maintained at $P_1$. At $T_B$ the pressure was then ramped down to $P_0$.

A variety of systems were studied, atomic, molecular and polymeric, four of which were known or found to be Roskilde liquids, and three that are not.

Atomic

- **Kob-Andersen Binary Lennard-Jones (KABLJ).** The well-studied KABLJ mixture, known to be a Roskilde liquid [1]; N=1000 particles.
- **Network Glass Former (NGF).** A network glass former that lacks *W-U* correlations and exhibits poor density scaling [33]; N=9000 particles.

Molecular

- **Asymmetric Dumbbell (AD).** Rigid asymmetric dumbbell that is Roskilde simple (see [34,35] for details); N=1000 molecules.
- **Short Asymmetric Dumbbell (SAD).** Same as the Asymmetric Dumbbell system but 20% shorter bond length; N=1000 molecules.
- **Asymmetric Dumbbell Mixture** (ADM)**.** Rigid asymmetric dumbbell (mixture) (see [36] for details) that has a prominent secondary relaxation. Each molecule is composed of two Lennard-Jones particle with size ratio $\sigma_B/\sigma_A$=0.625, connected by a rigid bond of length l=0.45; N=1000 molecules. This system differs from the two previous asymmetric dumbbell systems in that it is a Kob-Andersen-like 80:20 mixture in order to suppress crystallization; it also has stronger interactions between the smaller of the particles comprising the dumbbell and different particle size ratio and bond length.

Polymeric

- **Freely Jointed Chain (FJC).** Lennard-Jones freely-jointed chain; N=2000 particles (20 chains x 100 segments per chain). Non-bonded particles interact through LJ potential with σ=ε=1. Bonded particles connected by harmonic bonds with spring constant k=3000 and equilibrium bond length l=1.
- **Freely Rotating Chain with side group (FRC).** A more realistic, generic polymeric system (loosely based on polyisoprene) consisting of a freely-rotating polymer chain with a pendant group; N=8000 (16 chains x 500 segments). All bonded segments are connected by harmonic bonds with k=3000 and l=1. A harmonic bond angle potential with spring constant $k_a$=3000 and equilibrium angle 120° is applied to all bonds. Non-bonded main chain segments interact through LJ potential with $\sigma_{m\text{-}m}=\varepsilon_{m\text{-}m}=1$. Every fourth main chain segment is connected to a side group, an LJ particle with $\sigma_{s\text{-}s}=1.5$, $\varepsilon_{s\text{-}s}=1$. Cross-interactions are according to the Lorentz-Berthelot rules: $\sigma_{m\text{-}s}=1.25$, $\varepsilon_{m\text{-}s}=1$.

We quantify the amount of densification resulting from the temperature quench through $T_g$ by the parameter

$$\delta = \frac{v_N - v_D(P_0)}{v_N - v_D(P_1)} \tag{2}$$

where $v_N$, $v_D(P_0)$, and $v_D(P_1)$ are the respective specific volumes of the normal glass, the pressure densified glass, and the pressure densified glass prior to the removal of the pressure, all at the quench temperature $T_B$.

The pressure $P_1$ was chosen so that at $T_B$, the density was roughly 10% higher than for the conventional glass. The determination of whether a liquid is Roskilde simple for the atomic and rigid molecular systems was based on *W-U* correlations, as well as invariance of the radial distribution function at two state points at respective pressures $P_0$ and $P_1$ that have equal $\tau_\alpha$. For each system *W-U* correlations were evaluated from an NVT run at a single state point P~1 at a temperature for which $\tau_\alpha = 100$-1000.

## THEORETICAL BACKGROUND

To understand how the lack of pressure densification follows from isomorph theory, we consider the hypothetical cooling procedures depicted in Figure 1. For the conventional glass, starting at a state point L in the equilibrium liquid, we cool at pressure $P=P_0$, and at a constant cooling rate. At some point A the system falls out of equilibrium, and we continue to cool to a state point G far below the glass transition.

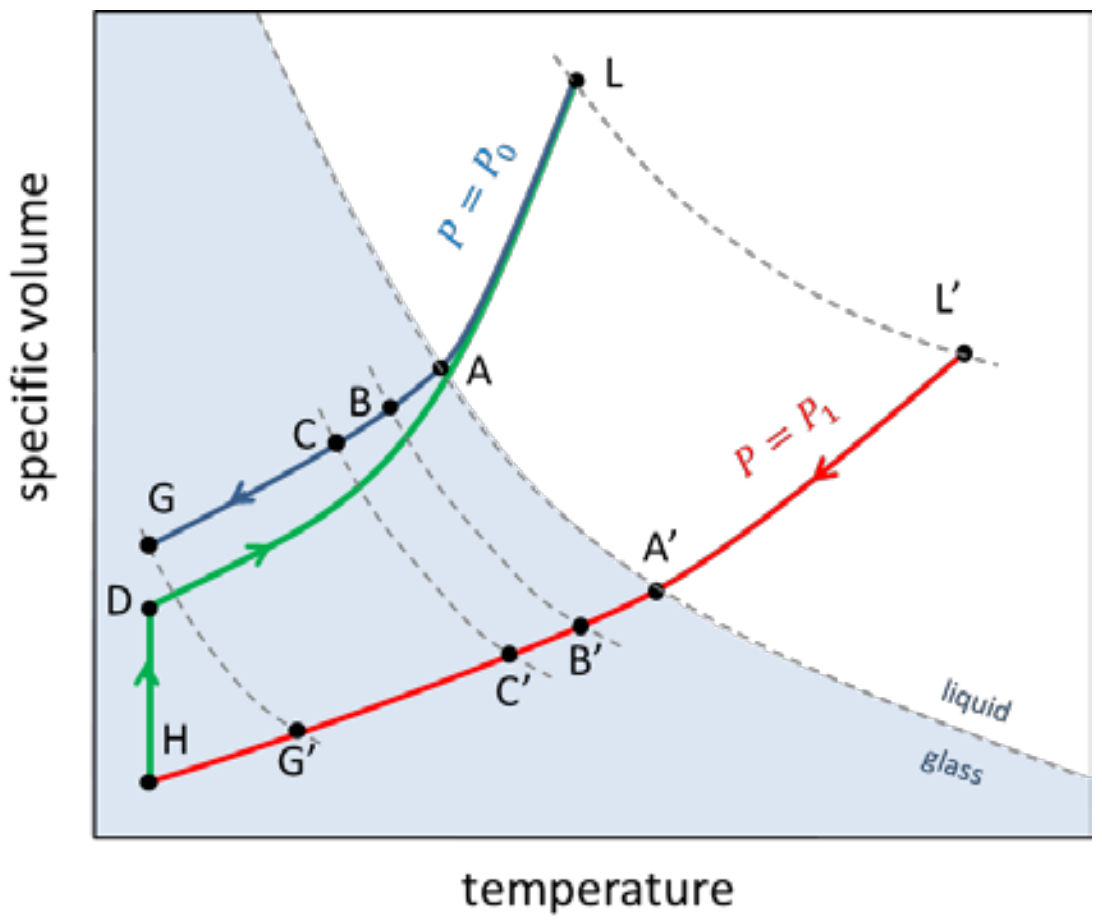

Figure 1. Schematic illustrating the prediction that liquids having isomorph behavior cannot be pressure densified. The dashed lines represent isomorphs in the equilibrium liquid, whereas in the glass they are calculated state points for which the *equilibrium* relaxation time is constant. In the glass, two systems falling on the same dashed line are not necessarily isomorphic because they could be in differently aged states.

For the pressure densified glass, starting at a point L' isomorphic to L, at a higher pressure, $P=P_1$, we adopt a cooling scheme L'→G' such that at every point in time the system is isomorphic to the first system moving from L to G. In the liquid state (L→A and L'→A') this is trivial, and one ends up at points A and A', which are on the "glassy isomorph". In the glass, keeping points on A'→G' isomorphic to those on A→G requires some care, since it's possible for two systems to be on the same dashed line in Fig. 1 but at different departures from equilibrium and thus not isomorphic. We break up each trajectory into a series of small steps. Consider a small time interval δt, during which the original system moves from point A to point B at the same pressure $P_0$, and slightly lower temperature and volume. For small enough δt, we can decompose this into a small instantaneous temperature and volume jump from A to B, followed by waiting at B for time δτ as the system relaxes from an initially slightly higher pressure to $P=P_0$. For the second system, starting at a point A' isomorphic to A, we *choose* a point B' isomorphic to B (on the same dotted line in Fig. 1) such that after an instantaneous jump A'-B', the system relaxes to pressure $P=P_1$ after the same time interval δt in reduced units. Thus, we have a jump from a pair of isomorphic state points (A, A') to another pair of isomorphic state points (B, B'). At B and B' respectively, the two systems have identical aging behavior; in fact the systems follow the same path in configuration space in reduced units and remain isomorphic. Continuing this process (from B, B' to C, C', etc), it is in principle possible to find a cooling protocol required to take the second system to a state point G' well below the glass transition, in a way that keeps it isomorphic to the first system at all times. Such cooling is not expected to be at a constant rate.

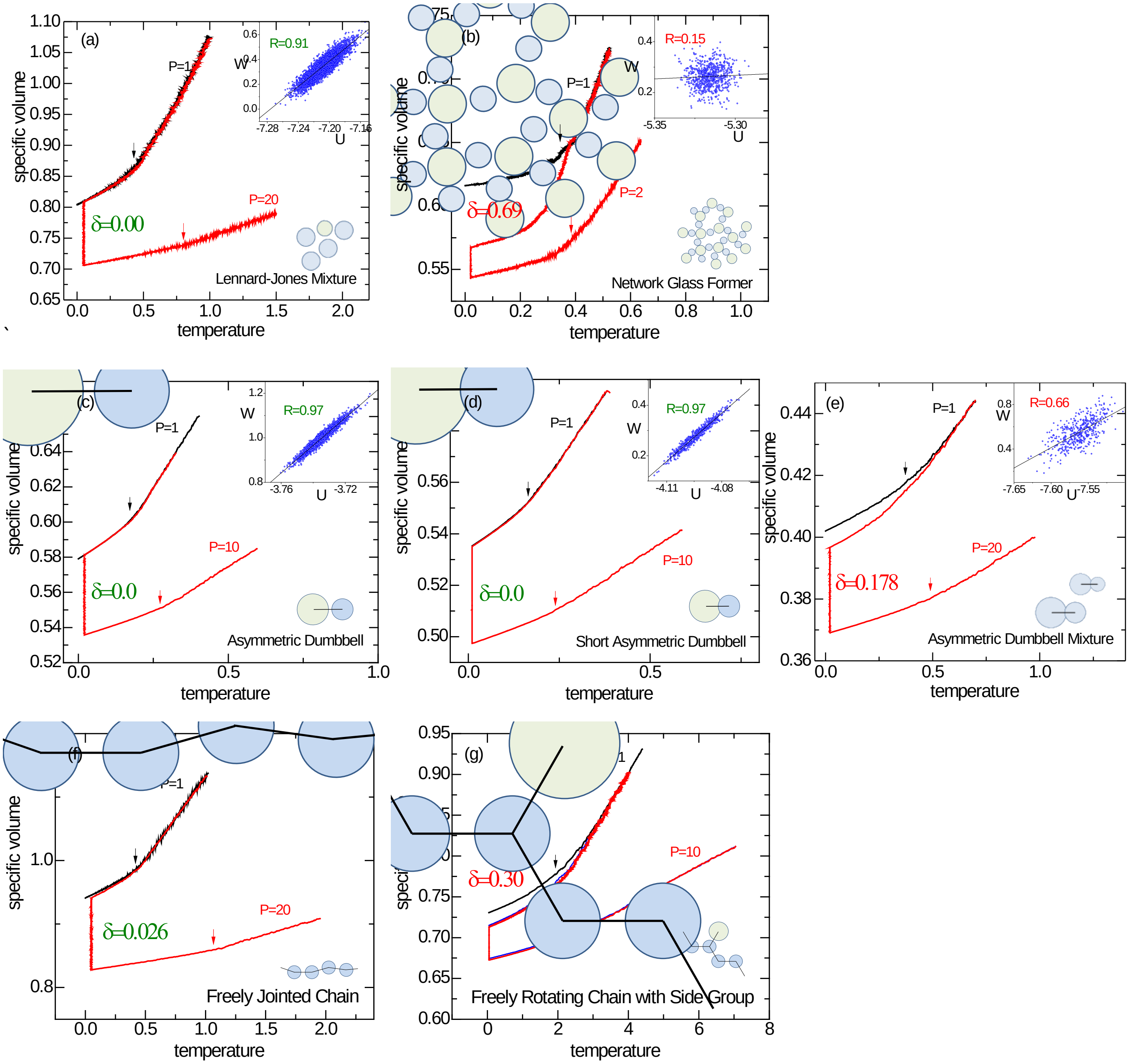

Figure 2. Specific volume as a function of temperature for the liquid cooled at low and high pressures, the later followed by release of the pressure. The Lennard-Jones Mixture, Asymmetric Dumbbell, Short Asymmetric Dumbbell and Freely Jointed Chain cannot be pressure densified. Inset shows the *W-U* correlation for the non-polymers.

The result of this hypothetical process is that if the second system (cooled at $P_1$) jumps from state point G' to G, it will be in an identical state as the first system (cooled at $P_0$) at G. If done sufficiently quickly (meaning, much faster than the aging rate), the jump from G' to G can be carried out along any path, for example that used in a typical pressure densification experiment: G' to a point H along a constant pressure path, and H to G at constant temperature. The last step, H→G, can equivalently be done as a pressure jump from $P_1$ to $P_0$ instead of a volume jump.

Table 1. Properties of systems studied herein

| | System | W-U correlation coefficient | RDF invariant at constant τ? | δ | secondary relaxation |
|---|---|---|---|---|---|
| atomic | Lennard-Jones Mixture (KABLJ) | 0.91 | yes | 0.00 | no |
| | Network Glass Former (NGF) | 0.10 | no | 0.69 | no |
| molecular | Asymmetric Dumbbell (AD) | 0.95 | yes | 0.00 | no |
| | Short Asymmetric Dumbbell (SAD) | 0.95 | yes | 0.00 | yes |
| | Asymmetric Dumbbell Mixture (ADM) | 0.66 | no | 0.18 | yes |
| polymeric | Freely Jointed Chain (FJC) | --- | yes | 0.03 | no |
| | Freely Rotating Chain with Side Group (FRC) | --- | no | 0.30 | yes |

Thus, isomorph theory predicts that pressure densification of a Roskilde liquid will "fail"; i.e., result in a glass identical to that cooled at low pressure. If a denser glass is obtained, the liquid is not Roskilde simple. From an experimental perspective, the high pressure cooling step needs to be done at a particular cooling rate, as described above. However, we find that the results are sensibly independent of cooling rate.

## RESULTS

Displayed in figures 2a-g for the 7 systems is the specific volume vs. temperature during (i) cooling at low pressure to form the normal glass, and (ii) cooling at elevated pressure with subsequent depressurization to form the densified glass.

For the KABLJ mixture (Fig. 2a), a prototypical Roskilde liquid [1], the densified glass is identical to the normal glass at the same temperature, and its heating scan overlaps that of the normal glass. On the other hand, the NGF (Fig. 2b), despite having only simple Lennard-Jones and inverse power law interactions, deviates significantly from isomorph theory [33]. It can be pressure densified: When the pressure is decreased to 1, the system recovers only 31% of the density difference from the normal glass.

The AD system (Fig. 2c) is a molecular liquid well known to be Roskilde simple [34,35], and it fails to pressure densify. On the other hand, the similar ADM liquid (Fig. 2e) has much weaker *W-U* correlations and a much larger difference in the intermolecular radial distribution function at equal relaxation times; it also shows significant pressure densification. The ADM liquid, unlike the AD system, has a prominent β relaxation. To assess whether the presence of a secondary relaxation influences the capacity for pressure densification, the SAD liquid was also tested (Fig. 2d). It is identical to AD but the bond length is 20% shorter, which gives rise to a secondary β relaxation at accessible time scales. Nevertheless, we find that SAD is a Roskilde liquid (strong pressure-energy correlation and invariance of structure at constant $\tau_\alpha$), and fails to pressure densify. Thus, the existence of isomorphs is unrelated to the presence of a secondary relaxation.

Finally we tested two polymeric systems. The FJC system lacks W-U correlations due to the flexibility of the backbone; thus, the term "pseudo-isomorphs" has been applied to it [37]. This refers to lines in the phase diagram along which intermolecular structure and dynamics are invariant. It (Fig. 2f) shows only a small degree of pressure densification; just 2.6% of the density difference is not recovered on removal of the pressure. On the other hand, in the FRC system the bond angles are constrained; there is also a pendant group. This system shows large deviations in structure at state points with equal $\tau_\alpha$, and thus lacks even pseudo-isomorphs. Consistent with the lack of isomorphs, it can be pressure densified (Fig. 1g).

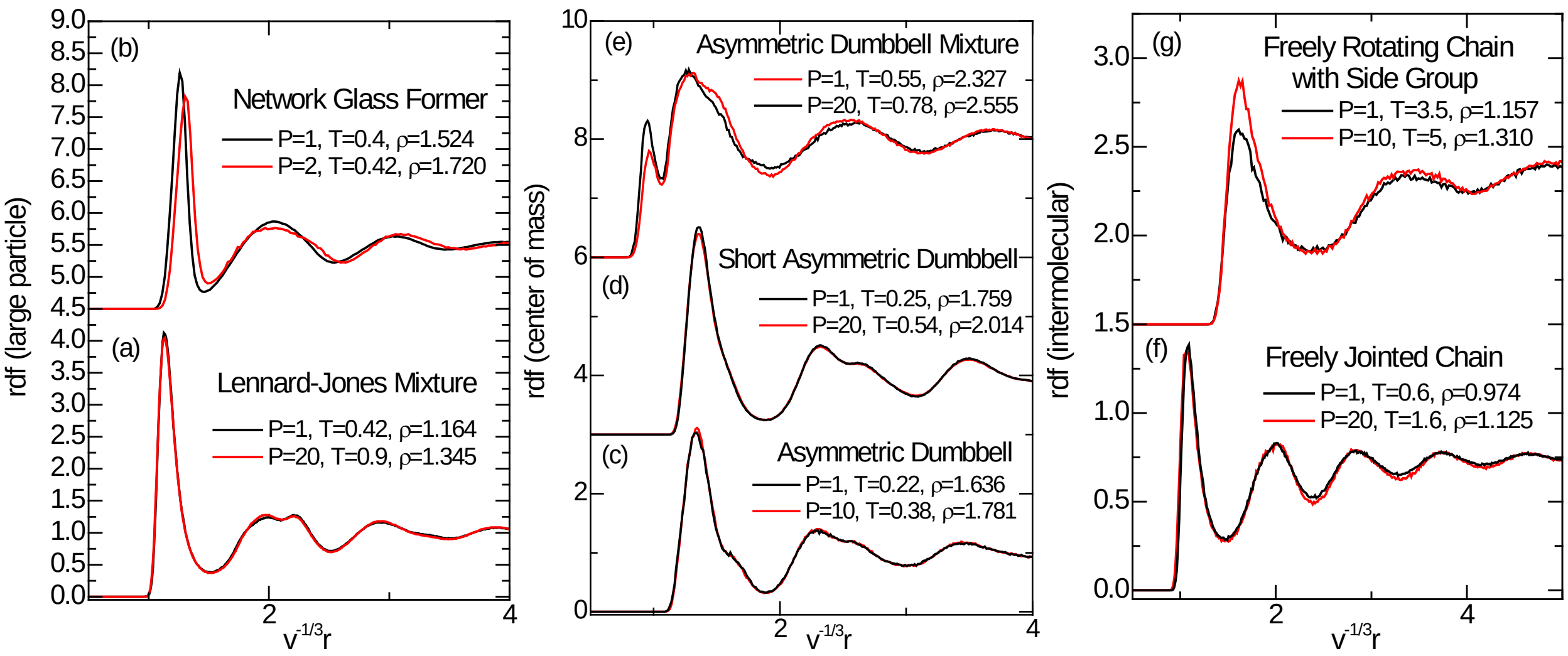

Figure 3. Radial distribution function for the seven systems studied herein. For each system the two state points shown have equal α relaxation time (in the range τ=100-1000 depending on the system) in reduced units. Only the systems which pressure densify exhibit isomorphs.

These seven systems show a correspondence between the existence of isomorphs and the absence of pressure densification. Furthermore, for the materials that can be pressure densified, the amount of pressure densification seems to correlate with the extent of departure from isomorphic behavior. This is a significant finding because we took no special care to control the cooling rate in order to remain on isomorphic state points during the low- and high-pressure cooling runs. To further demonstrate this insensitivity of the results to cooling rate, we repeated the pressure densification of the FRC system using a 10-fold faster rate of cooling. The result is included in Fig. 2g, where it is seen that the density is only slightly lower than for the more slowly cooled glass and much larger than the density of the normal glass.

From the definition of a Roskilde liquid in eq.(1) as having proportional probabilities for configurations with the same reduced coordinates, we expect this property for those materials herein that cannot be pressure densified. In figures 2a-g we compare for each liquid the radial distribution function at two state points associated with the same value of $\tau_\alpha$. The latter assures the state points are isomorphic, if the material has isomorphs. For the two atomic systems, the RDF is only shown for the larger particle. For the three molecular liquids, the RDF is for the center of mass, which is the one predicted to be invariant. For the polymers, only the intermolecular contribution to the RDF from the main chain atoms is computed. As can be seen, for the atomic (Fig. 3a) and molecular liquids (Fig. 3b), the equivalence of the RDF at state points having common $\tau_\alpha$ is found only for those materials that cannot be pressure densified. The presence of a secondary relaxation (SAD in Fig. 3b) does not affect the invariance of the RDF along an isomorph. For the polymers (Fig. 3c), the situation is less straight-forward due to the complication from intrachain motions. Considering only intermolecular bonds, the RDF is independent of state point for the FJC, which does not pressure densify, but shows significant differences between state points for the FRC, which shows significant pressure densification.

## DISCUSSION

The first preparation of glasses via pressure densification was by Tammann and Jenckel [21], with the technique having been applied to many materials: inorganic glasses such silica [28,29], hydrogen bonded liquids (phenolphthalein [21,22,38], sucrose [22], and glycerol [23]), the protic ionic liquid carvedilol dihydrogen phosphate [25], rosin (a mixture of organic acids) [21], and polymers including polyvinylethylene (PVE) [24], polystyrene (PS) [27,39,40], polymethylmethacrylate (PMMA) [41,42] and polyvinylchloride (PVC) [43]. In all cases it was reported that the glass cooled under pressure was denser, implying that none of these materials are Roskilde-simple. In some of the experimental studies of pressure densification the density of the compressed glass prior to the removal of pressure is reported, and an experimental δ parameter can be calculated: for PVC δ=0.15 [43], for atactic PMMA δ=0.24 [41], for polystyrene δ=0.13 [22], and for phenolphthalein δ=0.27 [38] (for these liquids $P_1$ was ~200-270 MPa and $P_0$ was ambient pressure).

With the exception of polymers, the ability of these materials to pressure densify is consistent with their reported properties: Network-forming glasses such as silica do not conform to density scaling [44] and their Prigogine-Defay ratio exceeds unity [8]. Hydrogen-bonded liquids and acids deviate from the behavior of Roskilde-liquids, due to the large effect that strong associations have on the state-point-dependence of the RDF [13,16].

The situation with polymers is less straight-forward. Density scaling has been demonstrated for PVE [15], PS [45], and PMMA [46], notwithstanding their capacity to be pressure densified. However, as seen in Figs. 1f and 1g, highly flexible chains seems to be reduce the capacity for pressure densification. That is, the freely-jointed chains exhibit the properties of Roskilde liquids, while the freely-rotating chains lack U-W correlation, but have some of the isomorph properties. None of the real polymers that have been pressure densified [24,27,39,40,41,42] are freely-jointed, so they are not necessarily Roskilde-simple.

## CONCLUSIONS

The simulation results are summarized in Table 1. Systems found to be Roskilde liquids, as evidenced by *U-W* correlation and a RDF that is invariant at fixed $\tau_\alpha$, do not pressure densify. Their density and thermal expansivity after vitrification under pressure are indistinguishable from those of glass produced conventionally by simple cooling. Significant deviation from isomorphic behavior, as reflected in a smaller correlation coefficient for *W-U* fluctuations and a RDF that varies with state point, was observed in those systems that could be pressure densified. The poorer the correlation of *W* and *U*, the greater the density difference between glass cooled at low and high pressure. The presence of isomorphs is not related to whether or not a material has a detectable secondary relaxation.

On the experimental side, all real materials tested to date exhibit pressure densification. Some of these, inorganic glasses and associated liquids, are known to lack isomorphs. However, polymers can also be pressure densified, even though they exhibit properties (e.g., density scaling and isochronal superpositioning) expected of materials having isomorphs.

## ACKNOWLEDGEMENTS

This work was supported by the Office of Naval Research. Stimulating discussions with R. Casalini and A. Holt are gratefully acknowledged.